\documentclass{ws-procs9x6-cpt19}
\begin{document}

\newcommand{\refeq}[1]{(\ref{#1})}
\def\etal {{\it et al.}}

\newcommand{\Pq}{q}
\newcommand{\Pu}{u}
\newcommand{\Rq}{R_{\Pq}}
\newcommand{\Ru}{R_{\Pu}}

\newcommand{\stwo}{S5 B0716+714}

\newcommand{\sourceindex}{}

\newcommand{\thetasphis}{\ensuremath{(\theta_{\sourceindex},\phi_{\sourceindex})}}

\newcommand{\Yjm}{\ensuremath{Y_{jm}}}
\newcommand{\Yjmi}{\ensuremath{Y_{jm{\sourceindex}}}}
\newcommand{\Yjma}{\ensuremath{\Yjm\thetasphis}}
\newcommand{\oYjm}{   \ensuremath{ {}_{0}\Yjm }    }
\newcommand{\oYjmi}{   \ensuremath{ {}_{0}\Yjmi }    }
\newcommand{\twoYjmi}{   \ensuremath{ {}_{\pm2}\Yjmi }    }
\newcommand{\kdVjm}{\ensuremath{\smash{k^{(d)}_{(V)jm}}}}

\newcommand{\kdV}{\ensuremath{\smash{\bar{k}^{(d)}_{(V){\sourceindex}}}}}
\newcommand{\kdVo}{\ensuremath{\smash{\bar{k}^{(d)}_{(V)}}}}

\newcommand{\kdVoo}{\ensuremath{\smash{k^{(d)}_{(V)00}}}}

\newcommand{\kdEB}{\ensuremath{\smash{\bar{k}^{(d)}_{(EB){\sourceindex}}}}}

\newcommand{\kdEBo}{\ensuremath{\smash{\bar{k}^{(d)}_{(EB)}}}}

\newcommand{\zi}{(z_{\sourceindex})}

\newcommand{\zz}{z^{\prime}}
\newcommand{\aaa}{a^{\prime}}

\newcommand{\zn}{\zi}

\newcommand{\Dpsi}{\Delta \psi^{(d)}\zn}
\newcommand{\Ldz}{\smash{L^{(d)}\zn}}
\newcommand{\Pqdz}{\smash{q^{(d)}\zn}}
\newcommand{\Pudz}{\smash{u^{(d)}\zn}}
\newcommand{\Pmaxdz}{\smash{p^{(d)}_{{\rm max}}\zn}}
\newcommand{\pmax}{p_{max}}

\newcommand{\pstar}{p_{\star}}
\newcommand{\sigmapstar}{\sigma_{\pstar}}
\newcommand{\sigmap}{\sigma_{p}}
\newcommand{\sigmaq}{\sigma_{q}}
\newcommand{\sigmau}{\sigma_{u}}

\newcommand{\apol}{APPOL}

\newcommand{\blazars}{Blazars}

\newcommand{\seq}{\,{=}\,}
\newcommand{\sgeq}{\,{\geq}\,}
\newcommand{\sleq}{\,{\leq}\,}
\newcommand{\ssim}{\,{\sim}\,}
\newcommand{\sgtrsim}{\,{\gtrsim}\,}
\newcommand{\slesssim}{\,{\lesssim}\,}
\newcommand{\splus}{\,{+}\,}
\newcommand{\simrange}[2]{$\sim$ ${#1}$--${#2}$}
\newcommand{\slesssimrange}[2]{$\lesssim {#1}$--${#2}$}
\newcommand{\sgtrsimrange}[2]{$\gtrsim {#1}$--${#2}$}

\title{Standard-Model Extension Constraints on Lorentz and CPT Violation 
From Optical Polarimetry of Active Galactic Nuclei}

\author{Andrew S.\ Friedman,$^{1}$ 
David Leon,$^{1}$
Roman Gerasimov,$^{1}$
Kevin D.\ Crowley,$^{1}$
Isaac Broudy,$^{1}$
Yash Melkani,$^{1}$
Walker Stevens,$^{1}$
Delwin Johnson,$^{1}$
Grant Teply,$^{1}$
David Tytler,$^{1}$
Brian G.\ Keating,$^{1}$ 
and Gary M.\ Cole$^2$}

\address{$^1$Center for Astrophysics and Space Sciences, University of California,\\ San Diego, La Jolla, CA 92093, USA}

\address{$^2$Western Nevada College, Carson City, NV 89703, USA}

\begin{abstract}
Vacuum birefringence from Lorentz and CPT violation 
in the Standard-Model Extension 
can be constrained 
using ground-based optical polarimetry of extragalactic sources. 
We describe results from a pilot program 
with an automated system 
that can perform simultaneous optical polarimetry 
in multiple passbands on different telescopes 
with an effective $0.45\,$m aperture.\cite{friedman19b}
Despite the limited collecting area, 
our polarization measurements of AGN 
using a wider effective optical passband 
than previous studies 
yielded individual line-of-sight constraints 
for Standard-Model Extension mass dimension $d=5$ operators 
within a factor of about one to ten 
of comparable broadband polarimetric bounds 
obtained using data from a $3.6\,$m telescope 
with roughly $64$ times the collecting area.\cite{kislat17}
Constraining more general anisotropic Standard-Model Extension 
coefficients at higher $d$ 
would require more AGN 
along different lines of sight.
This motivates a future dedicated ground-based, 
multi-band, 
optical polarimetry AGN survey 
with $\sgtrsim{1}\,$m-class telescopes, 
to obtain state-of-the-art anisotropic Standard-Model Extension
$d \seq 4,5,6$ constraints, 
while also using complementary archival polarimetry. 
This could happen more quickly and cost-effectively 
than via spectropolarimetry 
and long before more competitive constraints 
from space- or balloon-based x-ray/$\gamma$-ray polarization measurements.
\end{abstract}

\bodymatter

\phantom{}\vskip10pt\noindent

The Standard-Model Extension (SME)\cite{kostelecky09} 
allows for the violation of both Lorentz and CPT symmetry. 
Some SME coefficients 
predict vacuum birefringence, 
resulting in a wavelength-dependent rotation 
of the plane of linear polarization for photons. 
Such SME effects would increasingly depolarize light 
traveling over cosmological distances, 
with stronger observable effects 
predicted at higher redshifts and higher energies. 
Broadband polarimetry and spectropolarimetry 
of high-redshift extragalactic sources 
can thus be used to place increasingly sensitive astrophysical bounds 
on the SME.

Since SME effects can vary across the sky, 
one requires multiple measurements 
along different lines of sight 
to adequately constrain the most natural anisotropic SME models: 
the number $N(d)$ of distinct anisotropic vacuum-birefringent SME coefficients 
increases   
according to $N(d) = (d-1)^2$ 
and $N(d) = 2(d-1)^2 -8$ 
for odd and even mass dimension $d$, respectively.
Thus,
the mass dimensions
$d = 5,7,9,\ldots$ and $d = 4,6,8,10,\ldots$,
require respective polarization measurements along at least 
$N(d)=  16,36,64,\ldots$ 
or $10,42,90, 154,\ldots$ 
independent lines of sight 
for full coefficient coverage. 

Space- or balloon-based x-ray/$\gamma$-ray polarimetry 
can yield very sensitive line-of-sight constraints on the SME.\cite{kostelecky13} However, 
there are currently only about ten 
published x-ray/$\gamma$-ray polarization measurements of gamma-ray bursts (GRBs) 
that are not upper/lower limits.\cite{kostelecky13,wei19} 
By contrast, 
there are thousands of AGN 
with published broadband optical polarimetry 
and hundreds with spectropolarimetry.\cite{sluse05,smith09} 
At present, 
it is thus much more feasible 
to obtain wider sky coverage quickly, 
including many sources over a range of redshifts, 
by analyzing archival polarimetry 
of the most highly polarized AGN, 
including BL Lacs, 
Blazars, 
and highly polarized quasars.

Ultimately, 
advances in space- or balloon-based x-ray/$\gamma$-ray polarimetry 
of high-redshift transient GRBs 
could provide significantly stronger bounds on
anisotropic SME coefficients  
than optical AGN polarimetry in the coming decades.\cite{pearce19,kislat19,kostelecky13,kislat17}
However, 
since ground-based optical polarimetry and spectropolarimetry 
have smaller statistical and systematic errors 
than the more expensive and difficult x-ray/$\gamma$-ray polarimetry measurements, 
optical studies of AGN---the brightest, continuous, highly polarized, extragalactic optical sources---represent the most cost-effective approach 
to improve constraints on anisotropic SME coefficients today.\cite{kislat17,kislat18,friedman19b}

One can test CPT-odd birefringent SME coefficients 
with broadband polarimetry as follows.\cite{kostelecky09,kislat17,friedman19b} 
The $\smash{k^{(d)}_{(V) jm}}$ SME coefficients
predict a rotation of the linear polarization plane. 
For two photons with energies $E_1 < E_2$
emitted in the rest frame of a source at redshift $z$ 
with the same polarization,
the difference in polarization angle observed on Earth is
 \begin{eqnarray}
\Dpsi & \approx &
\left(E_2^{d-3}{\scriptstyle -}E_1^{d-3}\right) \Ldz \sum_{jm} \Yjma \kdVjm \, ,
\label{eq:deltapsi}
\end{eqnarray}
where $\Yjma$ are the spin-weighted spherical harmonics 
with celestial coordinates $\thetasphis$, 
and $\Ldz = \int_{0}^{z_{\sourceindex}} \frac{ (1+\zz)^{d-4} }{ H(\zz) }d\zz$ is the effective comoving distance;
$H(z)$ is the Hubble parameter for a FRW cosmology. 

If we conservatively assume 100\% intrinsic source polarization 
at all wavelengths,\cite{kislat17,kislat18,friedman19b}
integrating Eq.\ \refeq{eq:deltapsi}
over the effective energy bandpass $T(E)$ 
yields intensity-normalized Stokes parameters $\Pqdz$ and $\Pudz$ via
\begin{eqnarray}
\Pqdz + i\Pudz && = \int_{E_1}^{E_2}\!\! \exp \Big[ i \left(E^{d-3}-E_1^{d-3}\right)  \xi(z) \Big] T(E)\, dE \, ,
\label{eq:PqLIV}
\end{eqnarray}
where $\xi(z) \equiv 2\Ldz\,\kdV$ and $\kdV\equiv \sum_{jm} \Yjma \kdVjm$. 

The polarization
$\Pmaxdz = \big( [\Pqdz]^2 + [\Pudz]^2 \big)^{1/2}$ 
represents the theoretical maximum observable in the SME. 
Measuring a polarization fraction $\pstar \pm n\sigmapstar < \Pmaxdz$ 
can thus directly yield an $n$--$\sigma$ upper bound 
on the coefficient combination $\kdV$. 
A spherical-harmonic decomposition 
on the sky can then be used 
to combine sufficient numbers of line-of-sight constraints 
to bound all $N(d)$ parameters at a given $d$.\cite{kislat17,kislat18} 
These can also include line-of-sight constraints from spectropolarimetry, 
which can be about two or three orders of magnitude more sensitive 
than broadband polarimetry for $d = 5$.\cite{kislat17} 
Tests of CPT-even $d \seq 4$ SME coefficients are discussed in Ref.\ \refcite{kislat18}. 

The Array Photo Polarimeter (\apol)
is a pilot program 
with an automated small telescope system.\cite{friedman19b} 
It can conduct high-cadence, 
faint-object, 
optical polarimetry 
in multiple passbands 
with polarization-fraction statistical errors 
$\sigmap$ \slesssimrange{0.5}{1}\% 
for targets with visual band magnitude $V$ \slesssimrange{14}{15} mag, 
and systematic errors $\sigmap \ssim 0.04$\%. 
These are competitive with the best ground-based optical-telescope measurements. \apol{} is located at StarPhysics Observatory in Reno, Nevada 
at an elevation of $1585\,$m 
and serves as a test bed for future polarimeters 
that could be installed on $\sgtrsim{1}\,$m-class telescopes 
capable of observing much fainter AGN.\cite{friedman19b} 

\apol{} uses dual-beam inversion optical polarimetry 
with Savart plate analyzers 
rotated through a 
half-wave-plate image sequence;\cite{tinbergen05} 
it employs an automated telescope, 
filter, 
and instrument-control system 
with five co-located telescopes on two mounts.
We combined simultaneous polarimetry 
from two co-located Celestron 11- and 14-inch telescopes 
with an effective $17.8\,$inch ($0.45\,$m) telescope diameter 
with $Lum$ and $I_c$ filters 
into an effective optical bandpass 
with high transmission 
over the $\lambda\simeq 400$--$900\,$nm range 
of the two filters.\cite{friedman19b}
This yields more stringent SME bounds 
than either filter alone 
and achieves the effective collecting power of a larger telescope. 

Our initial \apol{} campaign observed two sources: 
BL Lacertae and \stwo{} 
at redshifts $z = 0.069$ and $z = 0.31$, 
respectively. 
This can only give SME line-of-sight constraints 
or bound the $\kdVoo$ isotropic CPT-odd SME coefficient.\cite{friedman19b}
Simultaneous optical polarimetry 
with our $Lum \splus I_c$ filter 
can yield SME line-of-sight constraints 
that are theoretically up to ten times ($d=5$) and 30 times ($d=6$)  
more sensitive 
than in the $I_c$ band alone. 
Despite our small effective $0.45\,$m aperture,
we achieved $d = 5$ line-of-sight constraints 
within a factor of up to ten in sensitivity 
compared to relevant constraints using broadband optical polarimetry 
with a $V$-band filter on a $3.6\,$m telescope 
with roughly 64 times the collecting area.\cite{kislat17,friedman19b}

Using archival optical polarimetry and spectropolarimetry 
for AGN (and GRB afterglows), 
there is a unique opportunity to test $k^{(d)}_{(V)}$ SME coefficients 
about one to two orders of magnitude better 
than previous work.\cite{kislat17,kislat18}  
While most archival optical data used a single filter, 
we conjecture 
that simultaneous observations in as few as two filters 
could constrain $k^{(d)}_{(V)}$ more cost-effectively 
than spectropolarimetry on $\sgtrsim 2\,$m-class telescopes. 
We are testing this now 
by performing simultaneous two-band optical polarimetry 
on roughly $10$--$20$ of the brightest, 
highly polarized, 
AGN 
that an upgraded $0.5\,$m \apol{} system can reasonably observe. 
This will enable design-feasibility studies 
for a ground-based, 
multi-band, 
optical polarimetry survey of high-redshift AGN 
with $\sgtrsim 1\,$m-class telescopes.
Finally, 
with thousands of sources over the sky, 
archival and new optical polarimetry could, 
for the first time, 
provide sufficient data to constrain not just individual $k^{(d)}_{(V)}$, 
but also a possible redshift dependence of any SME coefficients, 
potentially revealing time variation of the underlying fields
and elucidating the role of the associated new physics 
over cosmic history.

\section*{Acknowledgments}
ASF and BGK acknowledge support 
from NSF INSPIRE Award PHYS 1541160 and UCSD's Ax Center for Experimental Cosmology. 
DT is supported in part by NSF award AST1413568.

\end{document}